\documentclass[5p]{elsarticle}

\usepackage{lineno,hyperref}
\usepackage{graphicx}
\usepackage{color}
\usepackage{amsmath}
\usepackage{amsfonts}
\usepackage{aas_macros}

\newcommand{\review}[1]{#1}

\DeclareMathOperator*{\argmin}{arg\,min}
\DeclareMathOperator{\asinh}{asinh}

\newcommand{\angstrom}{\mbox{\normalfont\AA}}

\makeatletter
\newcommand{\verbatimfont}[1]{\renewcommand{\verbatim@font}{\ttfamily#1}}
\makeatother

\journal{Astronomy \& Computing}

\biboptions{authoryear}

\newcommand{\vect}[1]{\mathbf{#1}}
\newcommand{\matr}[1]{\mathbf{\mathrm{#1}}}

\begin{document}

\begin{frontmatter}

\title{Photo-z-SQL: integrated, flexible photometric redshift computation in a database}

%% or include affiliations in footnotes:
\author[ELTEaddress,JHUaddress]{R\'obert Beck\corref{mycorrespondingauthor}}
\cortext[mycorrespondingauthor]{Corresponding author}
\ead{beckrob23@caesar.elte.hu}

\author[ELTEaddress]{L\'aszl\'o Dobos}

\author[JHUaddress2,JHUaddress3,JHUaddress]{Tam\'as Budav\'ari}

\author[JHUaddress]{Alexander S. Szalay}

\author[ELTEaddress]{Istv\'an Csabai}

\address[ELTEaddress]{Department of Physics of Complex Systems, E\"{o}tv\"{o}s Lor\'{a}nd University, 1/A. P\'azm\'any P\'eter s\'et\'any, 1117 Budapest, Hungary}
\address[JHUaddress]{Department of Physics and Astronomy, The Johns Hopkins University, 3400 N Charles Street, Baltimore, MD 21218, USA}
\address[JHUaddress2]{Department of Applied Mathematics and Statistics, The Johns Hopkins University, 3400 N Charles Street, Baltimore, MD 21218, USA}
\address[JHUaddress3]{Department of Computer Science, The Johns Hopkins University, 3400 N Charles Street, Baltimore, MD 21218, USA}

\begin{abstract}
We present a flexible template-based photometric redshift estimation framework, implemented in C\#, that can be seamlessly integrated into a SQL database (or DB) server and executed on-demand in SQL. The DB integration eliminates the need to move large photometric datasets outside a database for redshift estimation, and utilizes the computational capabilities of DB hardware. The code is able to perform both maximum likelihood and Bayesian estimation, and can handle inputs of variable photometric filter sets and corresponding broad-band magnitudes. It is possible to take into account the full covariance matrix between filters, and filter zero points can be empirically calibrated using measurements with given redshifts. The list of spectral templates and the prior can be specified flexibly, and the expensive synthetic magnitude computations are done via lazy evaluation, coupled with a caching of results. Parallel execution is fully supported. For large upcoming photometric surveys such as the LSST, the ability to perform in-place photo-z calculation would be a significant advantage. Also, the efficient handling of variable filter sets is a necessity for heterogeneous databases, for example the Hubble Source Catalog, and for cross-match services such as SkyQuery. We illustrate the performance of our code on two reference photo-z \review{estimation testing} datasets, \review{and provide an analysis of execution time and scalability with respect to different configurations.} The code is available for download at \url{https://github.com/beckrob/Photo-z-SQL}.
\end{abstract}


\end{frontmatter}

%\linenumbers

\section{Introduction}
\label{sec:intro}

In recent years, photometric redshift (often abbreviated as photo-z) estimation has become a vital and widespread tool in the astronomical repertoire. Large amounts of photometric data are produced by sky surveys such as the Sloan Digital Sky Survey \citep[SDSS]{York2000, Eisenstein2011, Alam2015}, but spectroscopic measurements are more scarce due to the longer observing times required. Therefore it is important to accurately estimate the redshift -- and thus the distance -- of objects based just on their photometry.

There are two distinct approaches to photo-z estimation -- the empirical approach starts from a training set with known redshifts and uses a machine learning method to perform the estimation \citep{Connolly1995,Wang1998,Wadadekar2005,Csabai2007,Carliles2010,Gerdes2010,Brescia2014,Beck2016b}, while the template-based approach fits a spectral template using synthetic photometry, \review{i.e. by computing the broad-band magnitudes corresponding to known filter transmission curves and a known spectrum at a given redshift}. The former approach is generally more accurate \citep{Csabai2003}, but the latter does not require an extensive training set, and is thus more flexible.

Within the family of template-based photometric redshift estimation methods, there are two main branches: those that search for the maximum likelihood, best-fitting spectral energy distribution (SED) template and record the redshift of this single SED \citep{Bolzonella2000, Csabai2000, Arnouts2002, Ilbert2006}, and Bayesian methods that aim to reproduce the full posterior redshift probability distribution based on the observations \citep{Benitez2000, Coe2006, Brammer2008, Budavari2009}. The latter method provides a more detailed answer, but it is computationally more expensive, and it does need more input information in the form of a prior (or, with a flat prior, the peak of the probability distribution gives the same result as the maximum likelihood method).

Large photometric catalogs are most often stored in relational databases (e.g. \textit{SkyServer}\footnote{\url{http://skyserver.sdss.org/}} for SDSS), and cross-matching tools that connect such databases also use the relational model \citep{Dobos2012b,Budavari2013,Han2016}. It would be advantageous to have a photo-z implementation that can work directly on data in the servers, and on results from cross-matches, by integrating into the well-established SQL language.

There are various software \review{packages} available that utilize either the template-based or the empirical approach to perform photometric redshift estimation\footnote{See \url{http://www.sedfitting.org/} for a collection of links to photo-z codes.}. However, to the best of our knowledge, none of the public codes have the ability to perform the computations within the database itself. Thus, they require moving the photometric catalogs outside of the database, and the results also have to be loaded into the DB separately. This is a cumbersome process, especially when the amount of data is in the PB range, as will be the case with upcoming surveys such as the LSST \citep{Ivezic2008} and Pan-STARRS \citep{Tonry2012}. The problem is also acute when on-demand photo-z would be needed to quickly process cross-match results.

Additionally, existing template-based photometric redshift estimation codes use a predefined set of broad-band photometric filters, and compute synthetic magnitudes in all of these bands before the actual photo-z calculations. This approach is not ideal for heterogeneous observations, such as the Hubble Source Catalog \citep{Whitmore2016}, where each object can have a different corresponding filter set, and the total number of filters is much larger than the number of available filters for an object. Also, this issue can become especially prominent if we consider that filter transmission curves can change over time -- if this time-dependence is to be properly taken into account, precomputing synthetic magnitudes is simply not an option.

\review{Even with all these difficulties, the template-based approach is at least feasible in such situations, while building a training set for empirical estimation is almost impossible when the input filter set is not static, and not defined beforehand. Moreover, cross-match results are generally not extensive enough to provide complete training set coverage in redshift and color-magnitude space, and empirical methods usually do not perform well when substantial extrapolation is necessary \citep{Beck2017}.}

\review{Thus, we decided to adopt the template-based photo-z approach, and} sought to address the issues \review{of existing codes} with our own implementation, named Photo-z-SQL. \review{We apply a computational innovation to deal with variable filter sets -- we use lazy evaluation for synthetic magnitudes, computing them only when they are needed, and caching them to be reused.}

The code has been developed in the C\# language, which can be run stand-alone, but can also be integrated into Microsoft SQL Server as a set of user-defined functions. Microsoft SQL Server is a commercial DB server application, which has been chosen because it enables using a general-purpose computer language via CLR integration \citep[see][for another astronomy-related use case]{Dobos2011}, and also because it is used by e.g. the SDSS and Pan-STARRS collaborations. 

The database integration has the added benefit of utilizing the computational capabilities of DB servers \review{-- i.e. bringing the computations directly to the data --}, which often \review{have multiple processors specifically to handle computationally intensive tasks}. Parallel execution is \review{fully} supported \review{by our code}, taking advantage of modern multi-core systems. The objective-oriented power of C\# also leads to flexibility in defining and providing priors, templates and filters.

\section{Photometric redshift estimation methods}

Our code implements both the maximum likelihood and the Bayesian template-based photo-z method as separate functions, following the standard approaches in the literature \citep{Ilbert2006, Coe2006}. This section details the calculations performed by our algorithm.

\subsection{Maximum likelihood method}
\label{sec:ML}

The maximum likelihood method can be summarized with the formulae

\begin{multline}
\label{eq:chi2}
\chi^2 \left(z^{\prime}, t^{\prime}, m^{\prime}_{0}\right) = \\ \frac{1}{2} \sum_{ij} \left(m_i-\left(s_i\left(z^{\prime},t^{\prime}\right) - m^{\prime}_{0}\right)\right) C^{-1}_{ij} \left(m_j-\left(s_j\left(z^{\prime},t^{\prime}\right) - m^{\prime}_{0}\right)\right)
\end{multline}
\review{and}
\begin{equation}
\label{eq:argmin}
\left(z_{\mathrm{phot}}, t, m_{0}\right) = \argmin_{\left(z^{\prime}, t^{\prime}, m^{\prime}_{0}\right)} \chi^2 \left(z^{\prime}, t^{\prime}, m^{\prime}_{0}\right)\review{,}
\end{equation}
where $z^{\prime}$ and $t^{\prime}$ are the redshift and the template SED, $m^{\prime}_{0}$ is a factor that scales the total flux of the template spectrum, $\vect{m}$ is the measured broad-band magnitude vector of an object, $\vect{s}\left(z^{\prime},t^{\prime}\right)$ is the synthetic magnitude vector of template $t^{\prime}$ at redshift $z^{\prime}$, and finally, $\matr{C}$ is the covariance matrix of magnitude errors between filters.

The fitting process involves iterating over a set of redshifts and templates, but not over $m^{\prime}_{0}$, since for a given $z^{\prime}$ and $t^{\prime}$ the $m^{\prime}_{0}$ value that yields the lowest $\chi^2$ can be determined algebraically. $\left(z_{\mathrm{phot}}, t, m_{0}\right)$ denote the set of best-fitting parameters, corresponding to the lowest $\chi^2$ value found during the iteration. $z_{\mathrm{phot}}$ is the photometric redshift estimate.

Most applications assume that magnitude errors are uncorrelated, in addition to being Gaussian with a standard deviation matching the estimated measurement error for the given object. Under the uncorrelated assumption, the $\matr{C}$ covariance matrix takes the form
\begin{equation}
\label{eq:uncorrelated}
C_{ij}=\sigma^2_i\delta_{ij}\review{,}
\end{equation}
where $\mathbf{\sigma}$ is the estimated magnitude error vector of the object, and $\matr{\delta}$ is the Kroenecker delta. The expression for $\chi^2$ can be simplified under this assumption \review{to yield}
\begin{equation}
\label{eq:chi2simple}
\chi^2 \left(z^{\prime}, t^{\prime}, m^{\prime}_{0}\right) = \frac{1}{2} \sum_{i} \frac{\left(m_i-\left(s_i\left(z^{\prime},t^{\prime}\right) - m^{\prime}_{0}\right)\right)^2}{\sigma^2_i}\review{,}
\end{equation}
\review{but} in many cases \review{the} assumption may simply not hold \citep{Scranton2005}. For this reason, our code \review{implements} using the full covariance matrix \review{$\matr{C}$}, which, to our knowledge, is a missing feature in other public codes.

\subsection{Bayesian method}
\label{sec:bayesian}

Using the notation introduced in Sect.~\ref{sec:ML}, the Bayesian approach starts from the expression

\begin{multline}
\label{eq:bayes}
P\left(z,t,m_{0}|\vect{m},\matr{C}\right) = \\ \frac{P\left(z,t,m_{0}\right)P\left(\vect{m},\matr{C}|z,t,m_{0}\right)}{\int dz^{\prime} \int dt \int dm_{0} P\left(z^{\prime},t,m_{0}\right)P\left(\vect{m},\matr{C}|z^{\prime},t,m_{0}\right)}\review{,}
\end{multline}
\review{which specifies} the probability $P\left(z,t,m_{0}|\vect{m},\matr{C}\right)$ of a redshift and a template SED -- scaled to a flux -- given the data, i.e. the measured magnitudes and magnitude errors (optionally taking the covariance between filters into account). $P\left(z,t,m_{0}\right)$ is the prior probability of the given template SED at the given redshift. $P\left(\vect{m},\matr{C}|z,t,m_{0}\right)$ is the likelihood that the given template and redshift produce the data. The denominator contains a factor that normalizes the total probability to $1$, but in practice this is not computed, and instead the final probability distribution is normalized. We note that we use the $\int dt$ term as shorthand for integrating over the model parameters that describe a template SED, $t$ does not represent a real number.

The likelihood term can be formulated \review{as}

\begin{equation}
P\left(\vect{m},\matr{C}|z,t,m_{0}\right) = \exp\left(-\chi^2 \left(z, t, m_{0}\right)\right)\review{,}
\end{equation}
where $\chi^2$ is the expression defined in Eq.~\ref{eq:chi2}. Again, assuming uncorrelated magnitude errors, the simplified version in Eq.~\ref{eq:chi2simple} can be applied, but our application does support using the full covariance matrix \review{$\matr{C}$}.

The method aims to extract the posterior redshift probability distribution, $P\left(z|\vect{m},\matr{C}\right)$, which can be obtained by integrating out the nuisance parameters in Eq.~\ref{eq:bayes} \review{such that}
\begin{equation}
\label{eq:bayesfinal}
P\left(z|\vect{m},\matr{C}\right) = \int dt \int dm_{0} P\left(z,t,m_{0}|\vect{m},\matr{C}\right)\review{.}
\end{equation}

In the actual implementation, the integrals are discretized, replaced by a summation over a predetermined set of template parameters\review{,} a range in flux\review{, and the points of a predetermined redshift grid, to finally yield the expression}
\begin{multline}
\label{eq:bayesfinal2}
\review{P\left(z|\vect{m},\matr{C}\right) =} \\ \review{ \frac{\sum_t \sum_{m_{0}} P\left(z,t,m_{0}\right) \exp\left(-\chi^2 \left(z, t, m_{0}\right)\right)}{\sum_{z^{\prime}} \sum_t \sum_{m_{0}} P\left(z^{\prime},t,m_{0}\right) \exp\left(-\chi^2 \left(z^{\prime}, t, m_{0}\right)\right)}.}
\end{multline}

 The exact form of prior to use has to be chosen from the list currently available (see Sect.~\ref{sec:sqlinterface} and Sect.~\ref{sec:configurations}), but further priors could be easily added to the code. The output result is the $P\left(z|\vect{m},\matr{C}\right)$ posterior redshift probability distribution, normalized to an integral probability of $1$ on the given redshift grid.

\section{Database integration}

Microsoft SQL Server provides a unique opportunity among relational database management systems in its Common Language Runtime (CLR) integration feature. Compiled assemblies produced by any Common Language Infrastructure (CLI) programming language, which include C++, C\# and F\#, can be loaded into a DB on a server, thus granting access to code with arbitrarily complex functionality.

The main difficulty in SQL-CLR integration pertains to the SQL interface of the code, which has to translate between storage types of the programming language and the types available in SQL, and additionally it has to conform to the limitations of user-defined functions and user-defined stored procedures in SQL. Conversions between basic types such as integers, floating-point numbers and strings are fairly trivial, but more complex containers do need to be implemented on a case-by-case basis, relying on the binary storage types of SQL, or potentially on user-defined types. The publicly available \textit{SqlArray} library \citep{Dobos2011,Dobos2012c} is a good example of this process, providing variable-length array functionality in SQL.

Most algorithms need to store data, maintain a state between different execution steps. In a SQL interface, functions called in subsequent queries would normally only be able to share data by storing it on a disk, in DB tables using SQL types. However, in our implementation we bypass this limitation, and maintain a state in-memory without needing to convert to SQL types (see \review{Sect.~\ref{sec:sqlimpl}} below).

The Photo-z-SQL library has been coded in C\#. The interface consists of a set of user-defined SQL functions which can be used to first configure the photo-z calculation engine, and then perform the computations themselves. Parallel execution is fully supported to utilize the multi-core processors of typical DB servers. We provide install and uninstall scripts to make the installation process straightforward.

The remainder of this section gives more details about our implementation and interface in relation to the database integration.

\subsection{Implementation}
\label{sec:sqlimpl}
The data needs of the algorithm are the following: 

\begin{itemize}
	\item observed fluxes/magnitudes and errors for every object \citep[\review{optionally} with a][extinction map value]{Schlegel1998},
	\item the transmission curves corresponding to the broad-band filters,
	\item the spectral templates that are considered in the fit,
	\item information about the prior,
	\item additional configuration details such as the resolution in redshift and luminosity space.
\end{itemize}

In addition to the fluxes/magnitudes, the filter set may also change from object to object, but the rest of the data are static in a single estimation run, and thus can be pre-loaded and stored in the DB server. Additionally, the synthetic magnitude cache corresponding to the given configuration also has to be stored.

There is a way to perform this storage in-memory, without moving data into temporary tables: a static C\# class that uses the singleton pattern will retain its state in Microsoft SQL Server between function calls. Therefore we created a singleton wrapper class that stores the photo-z configuration and synthetic magnitudes. Currently only a single, global configuration can exist in a DB, \review{so multiple users would have to share the set of spectral templates, redshift grid and prior when estimating (but not the set of photometric filters, that can vary). In} the future this can easily be extended to allow separate configurations, e.g. linked to a \review{globally unique identifier (GUID) that corresponds to a} user.

The DB server needs access to filter and template curves. While this could be solved by locally loading the required data into tables, it would put an extra burden on users to create them, and to ensure that the photo-z code accesses the data correctly. Instead, we decided to use the Spectrum Services and Filter Profile Services of the Virtual Observatory (VO)\footnote{\url{http://voservices.net/}}. Users can choose from the existing templates and filters, or they can upload their own. This solution has the added benefit that the identifiers used in the Virtual Observatory can uniquely link the fluxes/magnitudes of objects to corresponding filters. The filter curves are also cached, thus they have to be downloaded from the VO only once.

The per-object data is expected to come from tables in the local DB server, supplied to the SQL functions of the Photo-z-SQL library. In order to allow variable-length array inputs of fluxes/magnitudes (and their errors), we used the \textit{SqlArray} library \citep{Dobos2011,Dobos2012c}, which can also be installed into a DB server via a simple script.

\begin{figure}
	\includegraphics[scale=0.22]{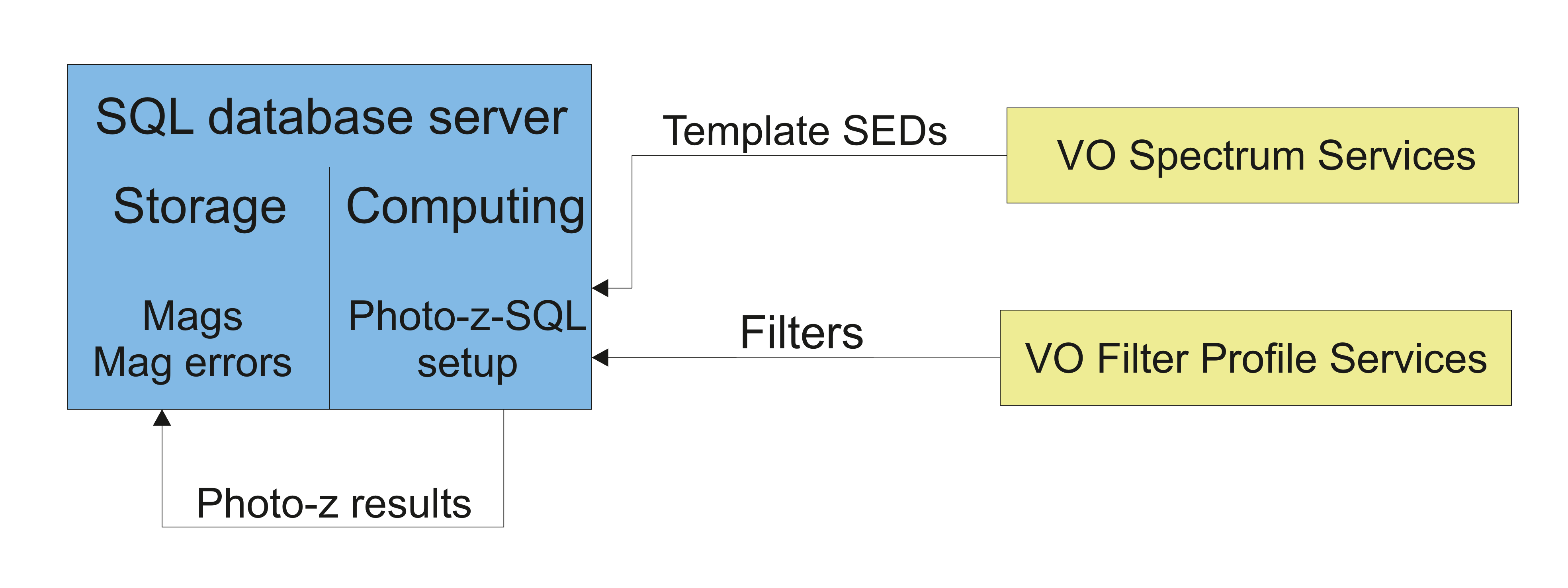}
	\caption{An illustration of the architecture of Photo-z-SQL. Computing refers to the dynamic memory and computing cores of the server, while storage refers to the disk-based data storage. Mag is shorthand for magnitude, but fluxes could also be used.}
	\label{fig:setup}
\end{figure}

The overall setup of Photo-z-SQL is illustrated in Fig.~\ref{fig:setup}.

\subsection{SQL interface}
\label{sec:sqlinterface}

Interaction with the Photo-z-SQL code is achieved with a collection of user-defined SQL functions that are called from the DB. There are \review{three} types of functions, Config functions that set up the photo-z configuration\review{,} Compute functions that perform the photo-z estimation itself\review{, and Util functions which e.g. help interfacing with \textit{SqlArray}.} When functions require a filter or template, they can be specified either with their VO integer identifier, or a complete URL address. These two options correspond to two versions of the given function, with a \_ID or \_URL suffix in the function name, respectively. Here we briefly list the main functions and their role, and an example query of a complete setup is provided in \ref{sec:queryexample}.

\begin{itemize}

\item \textbf{Config.SetupTemplateList} Specifies the list of SED templates to use in the photo-z estimation, along with the resolution in redshift and luminosity. Either the number of steps around the best-fitting luminosity can be given, or alternatively a range of physical luminosities (\_LuminositySpecified suffix).

\item \textbf{Config.SetupExtinctionLaw} Specifies the reference spectrum, $R_V$ dust parameter and extinction law to use when correcting for Galactic extinction (see Sect.~\ref{sec:extinction} for more details).

\item \textbf{Config.RemoveExtinctionLaw} Removes the data associated with the extinction law -- Galactic extinction correction is no longer applied.

\item \textbf{Config.SetupFlatPrior} Specifies that a flat prior should be used (this is the default prior).

\item \textbf{Config.SetupBenitezHDFPrior} Specifies that the HDF-N prior of \citet{Benitez2000} should be used (see Sect.~\ref{sec:configurations} for more details).

\item \textbf{Config.SetupAbsoluteMagnitudeLimitPrior} Specifies a maximum limit in absolute magnitude as the prior, in the provided reference filter. Cosmological parameters are also needed to define the distance--redshift relation.

\item \textbf{Config.SetupTemplateTypePrior} Specifies a prior that assigns a given probability to each of the used SED templates.

\item \review{\textbf{Config.AddMissingValueSpecifier} For convenience, specifies a value that denotes missing or otherwise invalid inputs. All such inputs are ignored.}

\item \textbf{Config.RemoveInitialization} Removes all data connected to the photo-z configuration, including cached filters, templates and synthetic magnitudes.

\item \textbf{Compute.PhotoZMinChiSqr} Performs maximum likelihood photo-z estimation, returning a single scalar $z_{\mathrm{phot}}$ value (see Sect.~\ref{sec:ML}). The prior is not used here, and luminosity is only evaluated at the best-fitting value. Magnitudes (or fluxes) and their errors, and corresponding filter identifiers are needed. An extinction map value can be specified, and the fit itself can be done in either magnitude or flux space. Also, an extra uncorrelated variance term can be added to the observational errors (see Sect.~\ref{sec:configurations}).

\item \textbf{Compute.PhotoZBayesian} Performs Bayesian photo-z estimation, returning the entire posterior redshift probability distribution within the specified coverage as a SQL table (see Sect.~\ref{sec:bayesian} for more details). Input parameters are the same as in Compute.PhotoZMinChiSqr.

\end{itemize}

\section{Implementation highlights}

In this section, we detail some additional the features of the implementation that could be noteworthy to users.

\subsection{Dynamic filter handling}

The computation of synthetic magnitudes is relatively expensive, as it involves integrating a filter curve with an SED curve. This has to be done -- ideally only once -- for every filter and SED pairing, at every considered redshift. Available photo-z software generally solve this issue by precomputing the entire synthetic magnitude table before the actual photo-z computations. As we noted in Sect.~\ref{sec:intro}, when the selection of filters changes between objects, or there is a large number of filters from which only a few are available at a time, or especially when filter time-dependence is to be taken into account, this choice is far from optimal.

In our implementation, we instead chose to set up a synthetic magnitude cache. The memory addresses of the filters serve as keys in a hash table, while the corresponding values are arrays of synthetic magnitudes. The arrays are indexed by the given template parameters and redshift. Whenever a synthetic magnitude is needed, it is retrieved from the cache if available, or computed and then stored in the cache if not. Therefore values are computed only when necessary, and only once.

With our approach, there is no requirement to know beforehand which filters will be needed. While it does include an additional hash table lookup whenever a synthetic magnitude is accessed, that is an inexpensive operation. Additionally, all functions working on the cache have been programmed to support parallel access. Currently, the cache has to be cleared by the user via a function call, but it would be possible to implement e.g. the deletion of old synthetic magnitudes after a time or cache size limit has been passed.

\subsection{Magnitudes and fluxes}

All of our photo-z algorithms have been programmed to be able to perform computations using either magnitudes or fluxes. When fluxes are used, the additive $m_{0}$ flux scaling parameter is replaced by a multiplicative $f_{0}$ parameter, but the best-fitting value of $f_{0}$ can still be determined algebraically for a given template SED and redshift (see Eq.~\ref{eq:chi2} and Sect.~\ref{sec:ML}).

Currently, our code supports using the AB magnitude system \citep{Oke1983}, and the SDSS $\asinh$ magnitude system \citep{Lupton1999}, but additional systems could be readily included. Functions are available for converting magnitudes and magnitude errors into fluxes and flux errors, and vice versa -- conversions between magnitude systems also use this mechanism. \review{The conversions assume a Gaussian distribution for errors, therefore users should take care not to perform such conversions when this assumption does not hold, e.g. in the case of faint objects where the flux errors are Gaussian, but the magnitude errors are not.}
 
\subsection{Zeropoint calibration and error scaling}
\label{sec:calibration}

Ideally, all photometric magnitude measurements should conform to the theoretical flux zeropoints of the given magnitude system, which, for example, is $3631\,Jy$ for the AB magnitude system \citep{Oke1983}. However, there are cases when this assumption is incorrect, and the actual zeropoint of a broad-band filter is slightly different \citep{Doi2010}.

To mitigate this issue, we implemented an optional zeropoint calibration algorithm which uses objects with known redshifts. The uncorrelated assumption is adopted here (see Eq.~\ref{eq:uncorrelated}), and the best-fitting template is determined at the given redshift using Eq.~\ref{eq:chi2simple} and Eq.~\ref{eq:argmin} for every object $j$ in the $T$ training set. Then, for each of the broad-band filters, indexed by $i$, a similar algebraic minimization is performed to find the filter-dependent zeropoint shift in magnitude, \review{expressed by}

\begin{equation}
m_i = \argmin_{\left(m^{\prime}_{i}\right)} \sum_{j \in T} \frac{\left(m_{i,j}-m^{\prime}_{i}-\left(s_i\left(z_j,t_j\right) - m_{0,j}\right)\right)^2}{\sigma^2_{i,j}}\review{.}
\end{equation}

Since the zeropoint shift can change the best-fitting template $t_j$ and the corresponding $m_{0,j}$, the two minimization steps are repeated iteratively -- with the latest $m_i$ shift applied to the $m_{i,j}$ measured magnitudes -- until a chosen zeropoint precision is achieved.

There is an additional, optional refinement to the calibration that can be applied after the iteration. If our assumptions are correct, and the data are well-described by the template spectra, the residuals scaled by the estimated photometric errors should follow the standard normal distribution. This will not be the case if the photometric errors are underestimated in a band, but the errors could be scaled by a factor
\begin{equation}
\alpha_i = \sqrt{\frac{1}{N_T-1}\sum_{j \in T} \frac{\left(m_{i,j}-m_i-\left(s_i\left(z_j,t_j\right) - m_{0,j}\right)\right)^2}{\sigma^2_{i,j}}}
\end{equation}
\review{to ensure that the scaled residual distribution}
\begin{equation}
\sigma^{\mathrm{scaled}}_{i,j} = \alpha_i \sigma_{i,j}
\end{equation}
\review{has a standard deviation of $1$. Above,} $N_T$ is the number of objects in the training set, and the error scaling factor is linked to a given filter \review{$i$}. Thus, problematic passbands with ill-estimated errors or a higher proportion of badly matching templates can be downweighted during the actual photo-z estimation.

We note that the same calibration algorithm can also be performed using fluxes, in that case the $f_i$ zeropoint flux multiplier is estimated for each filter.

\subsection{Galactic extinction}
\label{sec:extinction}

We provide a built-in implementation for correcting broad-band magnitudes for Galactic extinction using the IR dust map of \citet{Schlegel1998}. 

At present, two different extinction models are supported by the code. The first computes Eq.~B2 of \citet{Schlegel1998} and assumes the \citet{ODonnell1994} extinction law in the optical and the \citet{Cardelli1989} law in the UV and IR wavelength ranges. The second calculates Eq.~A1 of \citet{Schlafly2011} while assuming the \citet{Fitzpatrick1999} extinction law.

The reference galaxy spectrum used in the computation can be specified freely, but it should have coverage in the entire wavelength range of the broad-band filters for which the correction is applied.

\section{Photo-z results}

There have been two \review{recent} large publications \review{that allowed} the comparison of photometric redshift estimation methods \review{by publishing blind datasets for testing purposes.} \review{These two are} "Photo-z Accuracy Testing" or \textit{PHAT} by \citet{Hildebrandt2010}, and "A Critical Assessment of Photometric Redshift Methods: A CANDELS Investigation" by \citet{Dahlen2013}, which we will refer to as \textit{CAPR}. We use the public datasets of these articles to demonstrate the performance of our code.

\subsection{Configurations}
\label{sec:configurations}

We adopted some of the more successful approaches in the literature in our setups \citep{Hildebrandt2010, Dahlen2013}. We use two different sets of galaxy template SEDs, the Hubble UDF set of \citet{Coe2006}, denoted by \textit{BPZ}, and the COSMOS set of \citet{Ilbert2009}, indicated by \textit{LP}. In the case of the \textit{BPZ} set, following \citet{Coe2006} we linearly interpolated $9$ galaxies between each of the $8$ neighboring templates to generate a total of $71$ SEDs. The wavelength coverage of these templates ends at $25600 \angstrom$, after which they were linearly extrapolated. For the \textit{LP} set, adopting the choices of \citet{Ilbert2009} we used the \textit{Le PHARE} code \citep{Arnouts2002, Ilbert2006} to add emission lines of different fluxes, and to apply a selection of extinction laws with different parameters to the templates, yielding a total of $641$ SEDs.

The Bayesian estimation method was selected. The resolution of the redshift grid was taken to be $0.01$. We use two different priors, a simple flat prior (indicated by \textit{Flat}), and the apparent $I$-band magnitude, galaxy type and redshift prior of \citet{Benitez2000}, empirically calibrated on HDF-N data (denoted by \textit{HDF}). The latter prior has been adapted to the \textit{LP} template set based on galaxy type, distributing the total probability of a given type evenly among its instances. Since a measured $I$-band magnitude may not be available, the synthetic $I$-band magnitude corresponding to the given parameters is used as a proxy \review{to allow applying this prior in all situations}.

We found that adding an independent magnitude variance term to the measured magnitude variance can improve the estimation results. This extra error term can represent uncertainties in Galactic extinction, or in the template SEDs themselves, and it prevents single filters with very low errors from placing too stringent limits on the fitted templates. Thus, as an additional refinement, we may add an extra $0.01$ mag of error when using the \textit{LP} set, or $0.02$ mag in the case of the \textit{BPZ} set (shown by the tag \textit{Err}). The exact values were chosen based on the redshift estimation results of the \textit{PHAT} dataset, but have not been fine-tuned.

The combination of the two template sets, two priors, and whether or not extra error is added yields a total of $8$ different configurations that we present.

\subsection{PHAT}

\label{sec:phatresults}

The PHAT1 dataset \citep{Hildebrandt2010} contains $515$ galaxies that have a spectroscopic redshift, with magnitude measurements in $14$ different broad-band filters that span the wavelength range between $3000 \angstrom$ and $25000 \angstrom$ (here we do not use the extra $4$ IRAC filters). The measures that we report are the $\overline{\Delta z}_{\mathrm{norm}}$ bias and $\sigma \left(\Delta z_{\mathrm{norm}}\right)$ standard deviation -- excluding outliers -- of the normalized redshift error, $\Delta z_{\mathrm{norm}} = \frac{z_{\mathrm{spec}}-z_{\mathrm{phot}}}{1+z_{\mathrm{spec}}}$, and also the $P_o$ percentage of outliers. Outliers are defined as having $|\Delta z_{\mathrm{norm}}| > 0.15$. $z_{\mathrm{phot}}$ is chosen to be the highest-probability redshift in the posterior redshift distribution defined in Eq.~\ref{eq:bayesfinal2}.

\begin{figure*}
\begin{minipage}[t]{\textwidth}
	\includegraphics{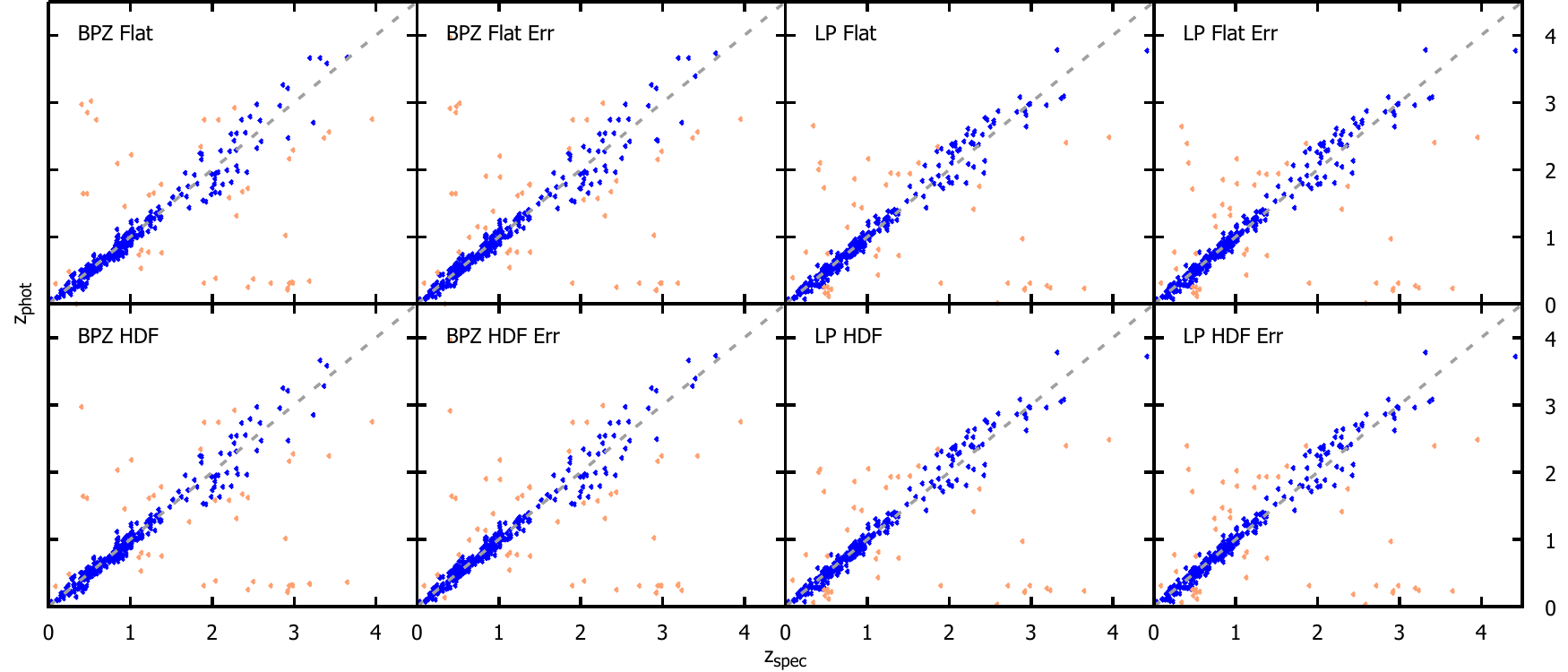}
	\caption{The $z_{\mathrm{phot}}$ photometric redshift as a function of the $z_{\mathrm{spec}}$ spectroscopic redshift, for all the configurations we ran when estimating the \textit{PHAT} dataset. The text in the top left corner of each panel indicates the given setup. Outlying galaxies with $|\Delta z_{\mathrm{norm}}| > 0.15$ are shown in \review{light} red, non-outlying galaxies in blue.}
	\label{fig:phatscatter}
\end{minipage}
\end{figure*}

The results are presented in Tab.~\ref{tab:phatresults}, while the spectroscopic--photometric redshift scatterplots are shown in Fig.~\ref{fig:phatscatter}. Applying the \textit{HDF} prior is slightly beneficial for the smaller \textit{BPZ} template set \review{in that it reduces scatter while only marginally changing other measures}, and \review{it is} somewhat detrimental in the case of the detailed \textit{LP} SED set\review{, mainly because of an increased bias}. Adding the extra error term reduces the estimation bias considerably while also slightly reducing the scatter, however, the outlier rate is marginally increased. All in all, our results are comparable to those of the better-performing methods in Table 5 of \citet{Hildebrandt2010}, whose typical values were: $\overline{\Delta z}_{\mathrm{norm}} \approx 0.004 - 0.0011$, $\sigma\left(\Delta z_{\mathrm{norm}}\right) \approx 0.038 - 0.048$ and $P_o \approx 9.2\% - 13.5\%$.

We note that the \textit{PHAT} dataset is not large enough to warrant performing the calibration detailed in Sect.~\ref{sec:calibration} \review{-- with such a small sample, the calibration generally converges to a local minimum in filter zeropoints (as opposed to the global one), improving performance on the calibration set, but actually decreasing it on the validation set.} Additionally, the \citet{Schlegel1998} dust map value corresponding to the galaxies was not published, \review{and neither were the coordinates on the sky,} therefore Galactic extinction has not been taken into account. 

\begin{table}
	\begin{tabular}{c | c | c | c}
		\hline
		
		Configuration & $\overline{\Delta z}_{\mathrm{norm}}$ & $\sigma\left(\Delta z_{\mathrm{norm}}\right)$  & $P_o$ \\
		
		\hline
		
		\textit{BPZ Flat} & 0.0114 & 0.0494 & 9.91\% \\

		\textit{BPZ HDF} & 0.0119 & 0.0484 & 9.94\% \\
		
		\textit{BPZ Flat Err} & 0.0029 & 0.0486 & 10.34\% \\
		
		\textit{BPZ HDF Err}  & 0.0026 & 0.0462 & 10.77\% \\
		
		\hline
		
		\textit{LP Flat} & 0.0049 & 0.0466 & 9.36\% \\
		
		\textit{LP HDF} & 0.0055 & 0.0467 & 9.38\% \\
		
		\textit{LP Flat Err} & 0.0013 & 0.0445 & 9.88\% \\
		
		\textit{LP HDF Err} & 0.0025 & 0.0448 & 10.08\% \\
		
		\hline
	\end{tabular}
	\caption{Numerical results for the \textit{PHAT} dataset. The \review{table lists the} $\overline{\Delta z}_{\mathrm{norm}}$ average bias and $\sigma\left(\Delta z_{\mathrm{norm}}\right)$ standard deviation of the $\Delta z_{\mathrm{norm}} = \frac{z_{\mathrm{spec}}-z_{\mathrm{phot}}}{1+z_{\mathrm{spec}}}$ normalized redshift estimation error, and the $P_o$ outlier rate with $|\Delta z_{\mathrm{norm}}| > 0.15$ outliers, for each configuration we ran.}
	\label{tab:phatresults}
\end{table}

\subsection{CAPR}

The \textit{CAPR} dataset \citep{Dahlen2013} contains $589$ galaxies intended for redshift estimation, along with a training set of $580$ galaxies intended for calibration. There are flux measurements in $14$ different passbands, of which we do not use the the IRAC $5.8 \mu m$ and $8.0 \mu m$ channels, similarly to code H (\textit{Le PHARE}) in \citet{Dahlen2013}. This way, the wavelength coverage is between $3000 \angstrom$ and $50000 \angstrom$. The IRAC passbands are problematic because they probe wavelength ranges where the spectral templates at low redshifts are not as reliable, and where Galactic extinction is a more significant factor \citep{Dahlen2013}. As with the \textit{PHAT} dataset, there were no published IR dust map values, therefore we did not take Galactic extinction into account.

\begin{figure*}
\begin{minipage}[t]{\textwidth}
	\includegraphics{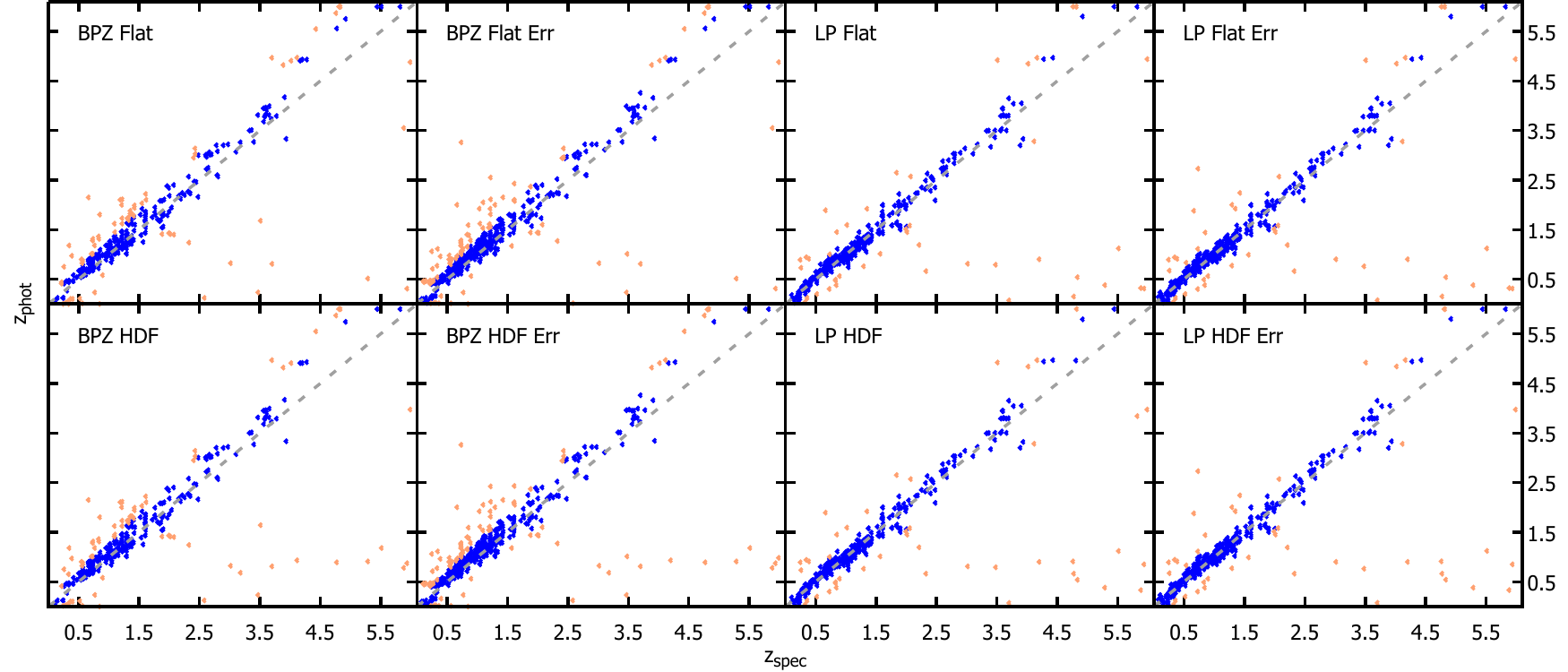}
	\caption{The $z_{\mathrm{phot}}$ photometric redshift as a function of the $z_{\mathrm{spec}}$ spectroscopic redshift, for all the configurations we ran when estimating the \textit{CAPR} dataset, without calibration. The text in the top left corner of each panel indicates the given setup. Outlying galaxies with $|\Delta z_{\mathrm{norm}}| > 0.15$ are shown in \review{light} red, non-outlying galaxies in blue.}
	\label{fig:caprscatter1}
\end{minipage}
\end{figure*}

First, our code was executed without performing any calibration. Again, we report the same numeric measures as in Sect.~\ref{sec:phatresults}, collated in Tab.~\ref{tab:caprresults1}. Refer to Fig.~\ref{fig:caprscatter1} for the redshift estimation scatterplots. Whether or not the \textit{HDF} prior was applied does not significantly impact the results, and the additional error term leads to a small improvement in the case of the \textit{LP} template set, but slightly worse bias and scatter for the \textit{BPZ} template set. \review{When compared with} the results in Table 2 of \citet{Dahlen2013} (bias, $\sigma_O$ and OLF columns), where typical values were: $\overline{\Delta z}_{\mathrm{norm}} \approx 0.005 - 0.023$, $\sigma\left(\Delta z_{\mathrm{norm}}\right) \approx 0.034 - 0.064$ and $P_o \approx 3.9\% - 9.3\%$, the \textit{BPZ} setups are among the worse performers, while the \textit{LP} configurations are in the middle of the pack. However, it should be noted that the better performers all included some sort of training using the spectroscopic sample.

\begin{table}
	\begin{tabular}{c | c | c | c}
		\hline
		
		Configuration & $\overline{\Delta z}_{\mathrm{norm}}$ & $\sigma\left(\Delta z_{\mathrm{norm}}\right)$  & $P_o$ \\
		
		\hline
		
		\textit{BPZ Flat} & -0.0359 & 0.0732 & 19.84\% \\

		\textit{BPZ HDF} & -0.0352 & 0.0727 & 20.49\% \\
		
		\textit{BPZ Flat Err} & -0.0428 & 0.0747 & 17.76\% \\
		
		\textit{BPZ HDF Err}  & -0.0421 & 0.0742 & 18.45\% \\
		
		\hline
		
		\textit{LP Flat} & -0.0083 & 0.0514 & 8.54\% \\
		
		\textit{LP HDF} & -0.0083 & 0.0514 & 8.54\% \\
		
		\textit{LP Flat Err} & -0.0083 & 0.0492 & 7.65\% \\
		
		\textit{LP HDF Err} & -0.0081 & 0.0490 & 7.65\% \\
		
		\hline
	\end{tabular}
	\caption{Numerical results for the \textit{CAPR} dataset, without calibration. The \review{table lists the} $\overline{\Delta z}_{\mathrm{norm}}$ average bias and $\sigma\left(\Delta z_{\mathrm{norm}}\right)$ standard deviation of the $\Delta z_{\mathrm{norm}} = \frac{z_{\mathrm{spec}}-z_{\mathrm{phot}}}{1+z_{\mathrm{spec}}}$ normalized redshift estimation error, and the $P_o$ outlier rate with $|\Delta z_{\mathrm{norm}}| > 0.15$ outliers, for each configuration we ran.}
	\label{tab:caprresults1}
\end{table}

\begin{figure}
	\includegraphics{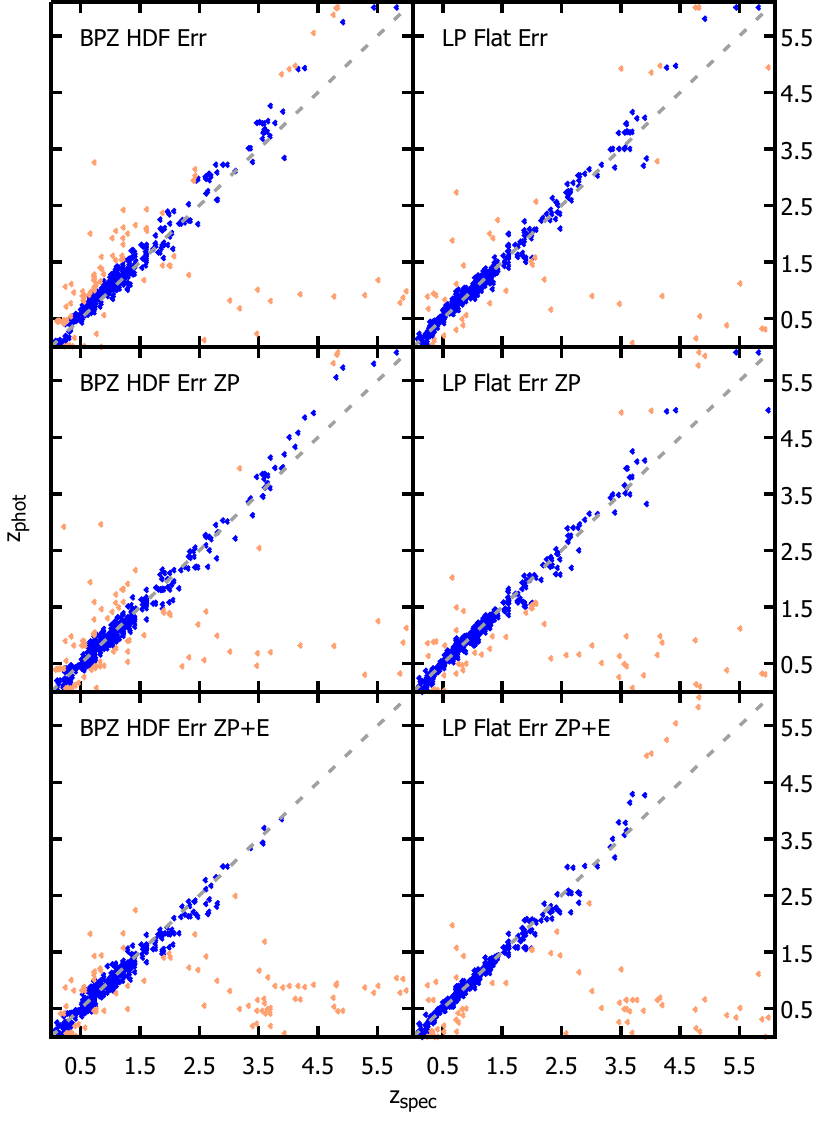}
	\caption{The $z_{\mathrm{phot}}$ photometric redshift as a function of the $z_{\mathrm{spec}}$ spectroscopic redshift, for the calibration tests we ran on the \textit{CAPR} dataset. The text in the top left corner of each panel indicates the given setup. Outlying galaxies with $|\Delta z_{\mathrm{norm}}| > 0.15$ are shown in \review{light} red, non-outlying galaxies in blue.}
	\label{fig:caprscatter2}
\end{figure}

For the sake of a better comparison, we tested our calibration algorithm, utilizing the published training set. We selected the \textit{BPZ HDF Err} and \textit{LP Flat Err} configurations to illustrate \review{the effects of calibration} -- the prior does not have a significant \review{influence on results}, and the added error was necessary because a few galaxies with very low estimated errors can adversely impact the calibration of a filter. As discussed in Sect.~\ref{sec:calibration}, we can either only perform zeropoint calibration (\textit{ZP}), or both zeropoint calibration and photometric error scaling (\textit{ZP+E}). The results are presented in Tab.~\ref{tab:caprresults2}, and the estimation scatterplots are shown in Fig.~\ref{fig:caprscatter2}. Even with the calibration, the \textit{BPZ} template set appears inadequate in describing the data, probably because its "proper" wavelength coverage ends at $25600 \angstrom$. However, after calibration the results of the \textit{LP} set are similar to what the better performers in \citet{Dahlen2013} can produce -- the only exception is the higher outlier fraction, which is dominated by high-redshift galaxies: $49\%$ of outliers have $z>3$, as opposed to only $8\%$ of the whole sample. In fact, the fraction of $z>3$ outliers jumps from $39\%$ to $52\%$ because of the zeropoint calibration, and to $76\%$ with the error scaling applied. This is because the training set is also mainly made up of low-redshift galaxies, and at different redshifts different passbands become important. Additionally, with low-redshift galaxies dominating the calibration, the less well-modeled higher wavelength ranges in the templates will lead to a larger than needed assigned uncertainty in the case of higher wavelength filters, downweighting filters that should constrain the photo-z estimation at high $z$. This issue could be solved by upweighting less well populated redshift bins in the calibration, and by applying a wavelength-dependent template error correction e.g. following \citet{Brammer2008}.

\begin{table}
	\begin{tabular}{c | c | c | c}
		\hline
		
		Configuration & $\overline{\Delta z}_{\mathrm{norm}}$ & $\sigma\left(\Delta z_{\mathrm{norm}}\right)$  & $P_o$ \\
		
		\hline
		
		\textit{BPZ HDF Err}  & -0.0421 & 0.0742 & 18.45\% \\		
		
		\textit{BPZ HDF Err ZP}  & 0.0117 & 0.0615 & 15.45\% \\		
		
		\textit{BPZ HDF Err ZP+E}  & 0.0200 & 0.0631 & 15.62\% \\		
		
		\hline
		
		\textit{LP Flat Err} & -0.0083 & 0.0492 & 7.65\% \\

		\textit{LP Flat Err ZP} & 0.0076 & 0.0445 & 9.85\% \\
		
		\textit{LP Flat Err ZP+E} & 0.0123 & 0.0402 & 12.05\% \\
		
		\hline
	\end{tabular}
	\caption{Numerical results for the \textit{CAPR} dataset, including calibration. The \review{table lists the} $\overline{\Delta z}_{\mathrm{norm}}$ average bias and $\sigma\left(\Delta z_{\mathrm{norm}}\right)$ standard deviation of the $\Delta z_{\mathrm{norm}} = \frac{z_{\mathrm{spec}}-z_{\mathrm{phot}}}{1+z_{\mathrm{spec}}}$ normalized redshift estimation error, and the $P_o$ outlier rate with $|\Delta z_{\mathrm{norm}}| > 0.15$ outliers. \textit{ZP} denotes only zeropoint calibration, while \textit{ZP+E} indicates zeropoint calibration and error scaling.}
	\label{tab:caprresults2}
\end{table}

\section{\review{Performance analysis}}

\label{sec:performance}

\review{In this section, we provide an analysis of the computational performance and memory usage of our code. We test how these factors scale when varying the details of the adopted configuration.}

\review{The test system was a slightly outdated server computer with a 16-core, 2.2 GHz Intel Xeon processor running Microsoft SQL Server 2012. With hyper-threading, this system can support 32 parallel threads.}

\review{The server could and was used by multiple users at the time, but no other queries were running that were as computationally intensive or lengthy as the Photo-z-SQL testing queries, leading to relatively stable benchmark results.}

\review{When first specifying the photo-z configuration, the synthetic magnitude cache is empty, and it will be filled during the starting phase of the first estimation query. Any subsequent queries that use the same filters will then work with the existing magnitude cache. For this reason, for each configuration we execute the same query twice, recording two times, $T_{\mathrm{first}}$ and $T_{\mathrm{second}}$. $T_{\mathrm{first}}$ includes the synthetic magnitude computation for the first few galaxies, the estimation time for the rest of the galaxies, and any overhead from the initial setup and the transmission of results. $T_{\mathrm{second}}$ includes the estimation time for all the galaxies, and again any overhead. Therefore $T_{\mathrm{cache}} \equiv T_{\mathrm{first}}-T_{\mathrm{second}}$ is a good measure of the time required to fill the magnitude cache, and using the number of galaxies $N_{\mathrm{gal}}$, $T_{\mathrm{gal}} \equiv T_{\mathrm{second}}/N_{\mathrm{gal}}$ is a reasonable estimate of the time required to perform photo-z on a single galaxy.}

\review{One of our goals is to quantify memory usage, however, C\# is a garbage-collected language with automatically managed memory, and it would be very difficult to track the memory used by different objects in a detailed way. Still, we can measure the total CLR memory usage $\mathrm{MEM}_{\mathrm{total}}$, which is the final metric in our analysis.}

\review{We use the \textit{PHAT} dataset in our tests, with $N_{\mathrm{gal}}=515$ (see Sect.~\ref{sec:phatresults}). Each test run was performed five times, and we provide the average and standard deviation of the resulting values. In Tab.~\ref{tab:performance}, we report the metrics defined above for the different configurations introduced in Sect.~\ref{sec:configurations}. We only mention the four \textit{Err} configurations, since the increased error values do not impact performance. As expected, having $\approx 9$ times as many template SEDs in the \textit{LP} configurations scales up runtime and memory usage by a correspondingly large factor of $\approx 8$. Additionally, applying the $HDF$ prior, which has to be evaluated in the innermost loop, more than doubles the per-galaxy estimation times. The synthetic magnitude computation time is not that much affected by the prior, since in that case the filter--spectrum integration is the most expensive operation.}
 
\begin{table*}
	\centering
	\begin{tabular}{c | c | c | c | c | c}
		\hline
		
		Configuration & $T_{\mathrm{first}}\left(s\right)$ & $T_{\mathrm{second}}\left(s\right)$ & $T_{\mathrm{cache}}\left(s\right)$ & $T_{\mathrm{gal}}\left(s\right)$ & $\mathrm{MEM}_{\mathrm{total}}\left(\mathrm{MB}\right)$\\
		
		\hline
		
		\textit{BPZ Flat Err} & $49.7 \pm 3.5$ & $18.1 \pm 1.1$ & $31.6 \pm 3.7$ & $0.035 \pm 0.002$ & $101 \pm 3$ \\
		
		\textit{BPZ HDF Err}  & $77.4 \pm 1.7$ & $41.4 \pm 2.8$ & $36.0 \pm 3.3$ & $0.080 \pm 0.005$ & $107 \pm 3$ \\
		
		\textit{LP Flat Err} & $296 \pm 29$ & $147 \pm 10$ & $148 \pm 31$ & $0.286 \pm 0.020$ & $760 \pm 19$ \\
		
		\textit{LP HDF Err} & $502 \pm 20$ & $348 \pm 5$ & $154 \pm 21$ & $0.676 \pm 0.010$ & $823 \pm 18$ \\
		
		\hline
	\end{tabular}
	\caption{\review{Performance results for four configurations on the \textit{PHAT} dataset. The table lists the execution time $T_{\mathrm{first}}$ of the Bayesian query for the first time (which includes synthetic magnitude computation), the execution time $T_{\mathrm{second}}$ of the same query for the second time (with synthetic magnitudes already cached), the cache filling time $T_{\mathrm{cache}}$, the photo-z estimation time $T_{\mathrm{gal}}$ for a single galaxy, and the total CLR memory usage $\mathrm{MEM}_{\mathrm{total}}$. See the text for a discussion.}}
	\label{tab:performance}
\end{table*}

\review{We follow up our tests of the predetermined configurations with an analysis of how the metrics scale when different parameters are varied. Starting from the \textit{LP Flat Err} configuration, again on the \textit{PHAT} dataset, we modify the number of templates used between $30$ and $630$, the number of broad-band filters considered between $4$ and the original $14$, the number of redshift grid points between $100$ and the original $600$, and finally the number of allowed parallel threads between $1$ and the original $32$.}

\begin{figure}
	\includegraphics{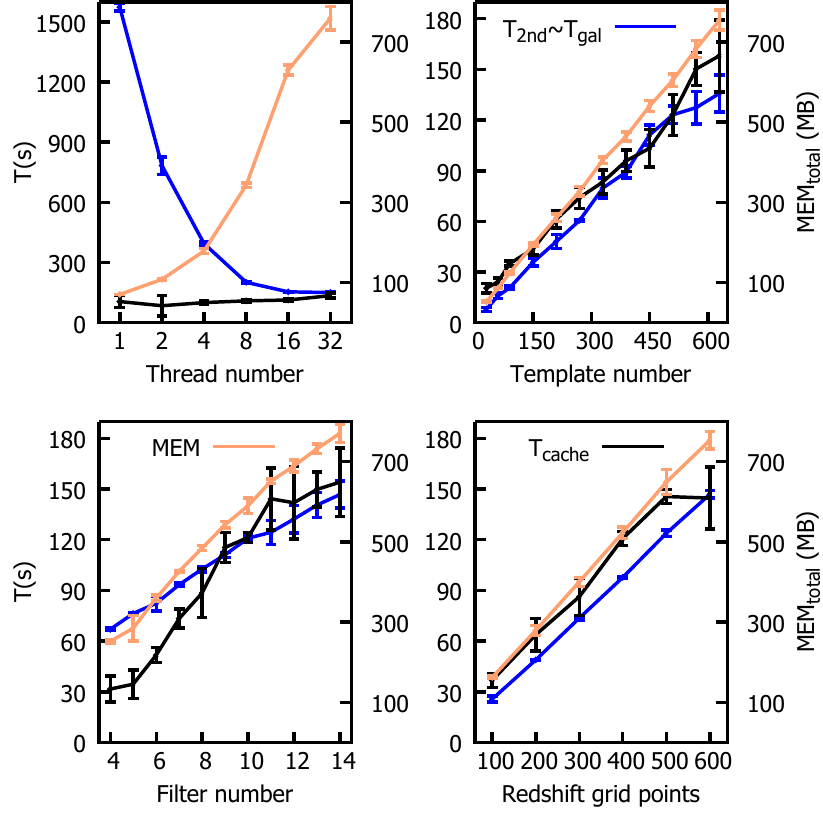}
	\caption{\review{The scaling of the execution time $T_{\mathrm{second}}$ (in blue, left axes), the cache filling time $T_{\mathrm{cache}}$ (black, left axes), and the total memory usage $\mathrm{MEM}_{\mathrm{total}}$ (light red, right axes) with respect to photo-z configuration parameters. To allow the information in $T_{\mathrm{cache}}$ and $T_{\mathrm{gal}}$ to share the same y axis, we show $T_{\mathrm{second}}$ in place of $T_{\mathrm{gal}}$, since $T_{\mathrm{second}}$ can be divided by $515$ to yield $T_{\mathrm{gal}}$. See the text for a discussion.}}
	\label{fig:performance}
\end{figure}

\review{Results from these scaling tests are presented in Fig.~\ref{fig:performance}. The most intriguing behavior can be observed in case of the thread number: the per-galaxy execution time $T_{\mathrm{gal}}$ drops sharply up to 16 threads as expected, but remains roughly the same for 32 threads. Since the number of cores in the server is $16$, this shows that hyper-threading is not an advantage for this algorithm. Also, the cache filling time $T_{\mathrm{cache}}$ remains roughly the same with an increasing thread number (in fact, it gets slightly worse due to more collisions), which is caused by the fact that all galaxies in our sample have the same filter list, filling the cache in the same order. However, this is a one-time cost, and could be parallelized if it ever becomes an issue. The allocated memory also sharply increases with more threads, which is expected since every thread requires resources to work on.}

\review{The execution time and memory usage functions with respect to redshift grid points, template number and filter number all follow the same, linearly growing trend. This also matches our expectations, since the storage unit sizes and number of required function evaluations scale linearly with these parameters. While the $T_{\mathrm{cache}}$ curve apparently breaks from the trend at filter numbers above $10$, this is due to the fact that those are infrared filters with a smaller resolution, hence the synthetic magnitudes are computed more quickly.}

\section{Summary}

In this article, we presented a new C\# photometric redshift estimation code, Photo-z-SQL. We detailed the photo-z approaches implemented by the code, and listed the most important available features. We demonstrated the performance of our code on two public datasets that had been used for photo-z method comparison, \textit{PHAT} and \textit{CAPR} \citep{Hildebrandt2010, Dahlen2013}. \review{We also provided an analysis of how execution time scales with various configuration parameter choices.}

Our photo-z estimation results are on par with those of the better performers in the literature, as expected from our adopted configurations, e.g. choice of template SEDs matching \citet{Ilbert2009}. \review{Interested readers are referred to \citet{Beck2017} for further method comparisons.} While we do not introduce anything inherently new from a photo-z estimation point of view, merely adopt some of the more successful methods in the literature, our implementation does possess the following important technical advantages:

\begin{itemize}
	\item Capability to be directly integrated into \review{Microsoft SQL Server}, eliminating the need to move photometric data outside the DB, and giving users and administrators an integrated platform for photo-z computations.

	\item Dynamic, on-demand handling of the set of broad-band photometric filters, which is a requirement for heterogeneous databases such as the Hubble Source Catalog \citep{Whitmore2016}.
	
	\item Ability to utilize the full covariance matrix between filters.
\end{itemize}

We should note that, while from a modeling standpoint it is better not to neglect the covariance between filters, in practice it may be difficult to appropriately determine the covariance matrix when no direct measurements are available. An empirical approach similar to the one in Sect.~\ref{sec:calibration} might be adopted, but that solution will depend on the adopted template set and calibration sample.

\review{Our stand-alone C\# code can be compiled and executed on Windows, Unix and Mac OS X systems, but the SQL-CLR integration is strictly tied to Microsoft SQL Server -- in this sense, the implementation is not portable. However, SDSS, Pan-STARRS and the Hubble Source Catalog all use Microsoft SQL Server, therefore there are plenty of useful applications even with this limitation.}

The implementation is actively in development, especially regarding priors, calibration techniques and the SQL interface. The planned applications include the assembly of a photo-z table for the Hubble Source Catalog, and integration into the SDSS \textit{SkyServer}, or the database cross-matching platform \textit{SkyQuery}\footnote{\url{http://www.sciserver.org/tools/skyquery/}} \citep{Dobos2012b,Budavari2013}.

The main unsolved challenge in running photo-z on heterogeneous, cross-matched data is the handling of different aperture sizes used in different catalogs. The differing apertures will result in a per-object magnitude shift between photometric bands, dependent on the apparent size of the object on the sky, and on what portion of it is covered by the apertures. Without going to the source images, these magnitude shifts cannot be transformed out -- unless there are multiple available aperture sizes for the same band, and some form of profile fitting can be utilized. More research will be needed on how strongly this issue affects photo-z results, and on how to mitigate the effects.

Another avenue for continuing this research could be the implementation of an empirical photo-z technique in C\#, to be integrated into Microsoft SQL Server. While the interface for specifying the training set and the learning step in SQL would presumably be more difficult to implement and to use than the interface of the template-based method, in principle there are no technical obstacles since a pre-loaded state can be stored in a static C\# class. Furthermore, the 2016 version of Microsoft SQL Server will provide support for R, potentially making it straightforward to perform empirical photo-z calculations directly in a DB.

The Photo-z-SQL code is available for download at \url{https://github.com/beckrob/Photo-z-SQL}.

\section*{Acknowledgments}

The realization of this work was supported by the Hungarian NKFI NN grant 114560. RB was \includegraphics[scale=0.045]{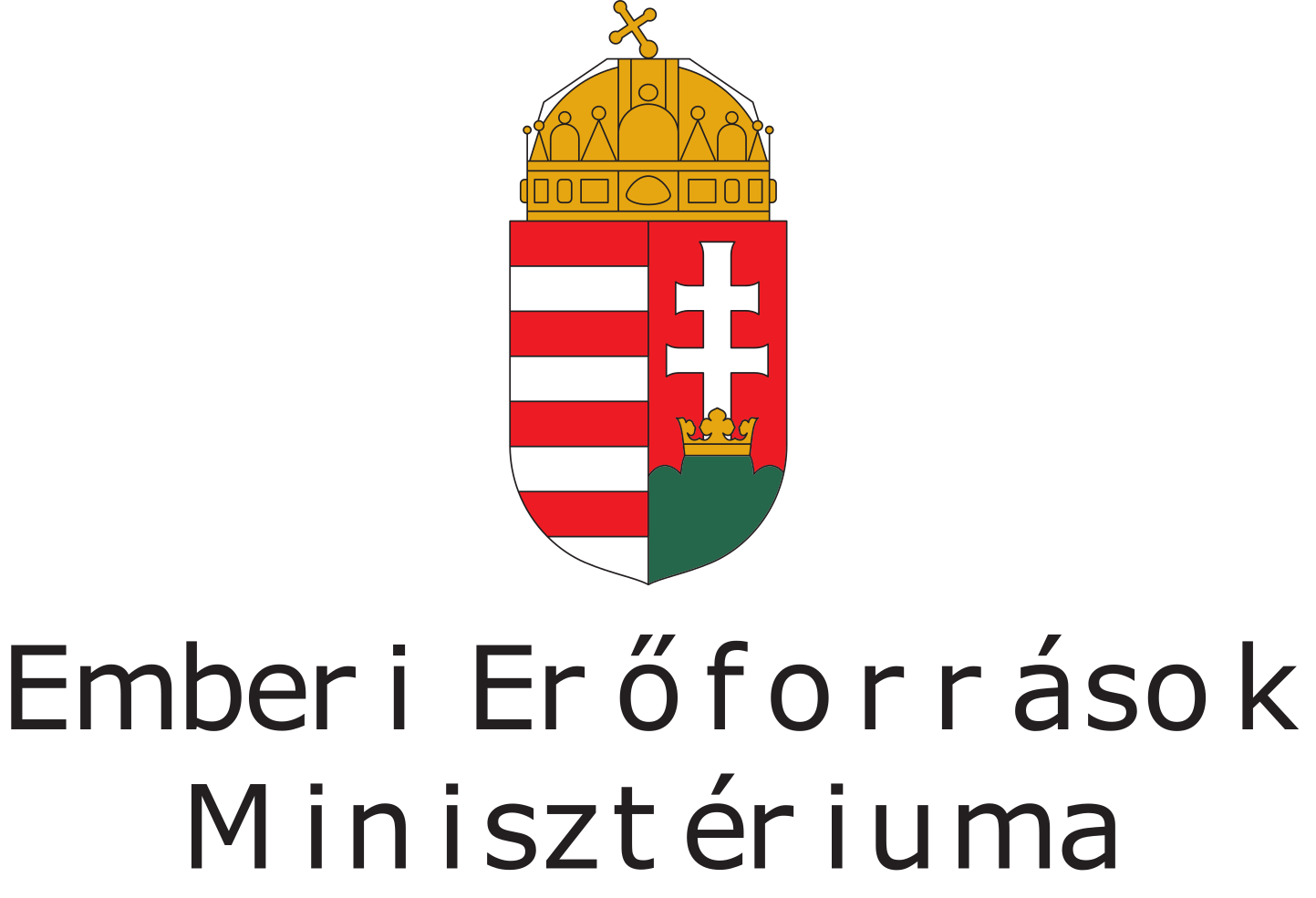} supported through the New National Excellence Program of the Ministry of Human Capacities, Hungary.

%\bibliography{photozsql_ac}

\vfill

\appendix

\section{Example SQL queries}
\label{sec:queryexample}

The first example sets up a photo-z configuration. First it clears the potentially existing photo-z setup, specifies that $-9999$ in the input data denotes a missing value, and chooses a flat prior. Then, it creates a string with VO template identifiers corresponding to the \textit{LP} template set (see Sect.~\ref{sec:configurations}), and parses it into a \textit{SqlArray} object. Finally, this template list is given to the photo-z code, with a redshift coverage between $0.001-1.001$ and a linear step size of $0.01$. $11$ steps will be taken in luminosity space around the best-fitting luminosity.

\verbatimfont{\footnotesize}
\begin{verbatim}
SELECT PhotoZSQL.Config.RemoveInitialization()

SELECT PhotoZSQL.Config.AddMissingValueSpecifier(-9999)

SELECT PhotoZSQL.Config.SetupFlatPrior()

--To get a SqlArray int array of template IDs, a string
--has to be set up with this format: '[511,512,(...),1151]'
DECLARE @IDString varchar(max) = '['
DECLARE @id INT = 511
WHILE @id < 1151
BEGIN
   SET @IDString = @IDString + CAST(@id AS varchar(max)) + ','
   SET @id = @id + 1
END
SET @IDString = @IDString + CAST(@id AS varchar(max)) + ']'

--Parsing the string into an integer array
DECLARE @TemplateIDArray varbinary(max) =
                SqlArray.IntArrayMax.ParseInvariant(@IDString)

SELECT PhotoZSQL.Config.SetupTemplateList_ID(@TemplateIDArray,
                                    1.0,0.001,1.001,0.01,0,11)
\end{verbatim}

The second example performs maximum likelihood photo-z estimation on a sample of SDSS data, using the $5$ SDSS filters. The fit is performed in magnitude space, with no additional extinction correction, and with a $0.01$ mag independent error term added.

\begin{verbatim}
DECLARE @FilterIDArray varbinary(max) = 
                 SqlArray.IntArrayMax.Vector_5(14,15,16,17,18)

SELECT TOP 100 gal.objID,
      PhotoZSQL.[Compute].PhotoZMinChiSqr_ID(
           SqlArray.FloatArrayMax.Vector_5(gal.dered_u,
                                           gal.dered_g,
                                           gal.dered_r,
                                           gal.dered_i,
                                           gal.dered_z),
           SqlArray.FloatArrayMax.Vector_5(gal.modelMagErr_u,
                                           gal.modelMagErr_g,
                                           gal.modelMagErr_r,
                                           gal.modelMagErr_i,
                                           gal.modelMagErr_z),
           1,@FilterIDArray,0.0,0,0.01) AS zphot
FROM DR12.Galaxy AS gal
\end{verbatim}

The third example performs Bayesian photo-z estimation, otherwise with the same particulars as the previous query.

\begin{verbatim}
DECLARE @FilterIDArray varbinary(max) = 
                 SqlArray.IntArrayMax.Vector_5(14,15,16,17,18)

SELECT TOP 100 gal.objID,pz.*
FROM DR12.Galaxy AS gal
CROSS APPLY PhotoZSQL.[Compute].PhotoZBayesian_ID(
           SqlArray.FloatArrayMax.Vector_5(gal.dered_u,
                                           gal.dered_g,
                                           gal.dered_r,
                                           gal.dered_i,
                                           gal.dered_z),
           SqlArray.FloatArrayMax.Vector_5(gal.modelMagErr_u,
                                           gal.modelMagErr_g,
                                           gal.modelMagErr_r,
                                           gal.modelMagErr_i,
                                           gal.modelMagErr_z),
           1,@FilterIDArray,0.0,0,0.01) AS pz

\end{verbatim}

\end{document}